\documentclass[fleqn,usenatbib]{mnras}

\usepackage{newtxtext,newtxmath}

\usepackage[T1]{fontenc}

\DeclareRobustCommand{\VAN}[3]{#2}
\let\VANthebibliography\thebibliography
\def\thebibliography{\DeclareRobustCommand{\VAN}[3]{##3}\VANthebibliography}

\usepackage{graphicx}	
\usepackage{amsmath}	
\usepackage{tabularray} 
\usepackage{bm} 		
\usepackage[normalem]{ulem} 
\usepackage[dvipsnames]{xcolor} 

\definecolor{dark}{HTML}{1e293b}

\title[Comparing dark matter and MOND]{Comparing dark matter and MOND hyphotheses 
from the distribution function of A, F, early-G stars in the solar neighbourhood}

\author[M. A. Syaifudin et al.]{
M. A. Syaifudin,$^{1}$\thanks{E-mail: \href{mailto:muhammad.ali.syaifudin@hotmail.com}{muhammad.ali.syaifudin@hotmail.com}}
M. I. Arifyanto,$^{2, 3}$\thanks{E-mail: \href{mailto:mochamad.ikbal@itb.ac.id}{mochamad.ikbal@itb.ac.id}}
H. R. T. Wulandari$^{2, 3}$,
F. A. M. Mulki$^{2, 3}$
\\
$^{1}$ Astronomy Master Program, Bandung Institute of Technology, Jl. Ganesa No. 10, 40135, Indonesia, \\
$^{2}$ Department of Astronomy, Bandung Institute of Technology, Jl. Ganesa No. 10, 40135, Indonesia, \\
$^{3}$ Bosscha Observatory, Bandung Institute of Technology, Jl. Peneropongan Bintang No. 1, 40391, Indonesia, \\
}

\date{Accepted 2024 October 6. Received 2024 October 6; in original form 2024 January 4}

\pubyear{2024}

\begin{document}

\label{firstpage}
\pagerange{\pageref{firstpage}--\pageref{lastpage}}
\maketitle

\begin{abstract}
	Dark matter is hypothetical matter assumed to 
	address the historically known as missing mass
	problem in galaxies. However, alternative theories, such as Modified
	Newtonian Dynamics (MOND), have been notably successful in explaining the
	missing mass problem in various astrophysical systems.
	The vertical distribution function of stars in the solar neighbourhood
	serves as a proxy to constrain galactic dynamics in accordance to its
	contents. We employ both the vertical positional
	and velocity distribution of stars in cylindrical coordinates with a radius
	of 150 pc and a half-height of 200 pc from the galactic plane. Our tracers
	consist of main-sequence A, F, and early-G stars from the GAIA, RAVE, APOGEE,
	GALAH, and LAMOST catalogues.
	We attempt to solve the missing mass in the solar neighbourhood,
	interpreting it as either dark matter or MOND. Subsequently, we
	compare both hypotheses newtonian gravity with dark matter and MOND,
	using the Bayes factor (BF) to determine which one is more favoured by the data.
	We found that the inferred dark matter in the solar neighbourhood is
	in range of $\sim (0.01 \textup{--} 0.07) \, \textrm{M}_{\odot} \, \textrm{pc}^{-3}$.
	The wide range of inferred dark matter density is caused by the peculiar behaviour of F-type stars, 
	which could be a sign of dynamical disequilibrium.
	We also determine that the MOND hypothesis's acceleration parameter $a_0$ is
	$(1.26 \pm 0.13) \times 10^{-10} \, \textrm{m} \, \textrm{s}^{-2}$ for simple
	interpolating
	function. The average of Bayes factor for all tracers between the two hypotheses is
	$\log \textrm{BF}\sim 0.1$, 
	meaning no strong evidence in favour of either the dark matter or MOND hypotheses.
\end{abstract}



\section{INTRODUCTION} \label{sec:introduction}

The study of dark matter constitutes a crucial topic in modern astrophysics.
The discovery of flat rotation curves at large radii in spiral galaxies
offered a glimpse into the existence of dark matter (Rubin et al. \citeyear{rubin1980rotational}).
The local density of dark matter, typically observed a few hundred parsecs from
the Sun, holds significant importance in constraining the nature of dark matter
within the Milky Way. This value then imparts information about the shape of the
Milky Way's dark matter halo in the midplane. Furthermore, it serves as a
constraint in galaxy formation models and cosmology (Read \citeyear{read2014local}), and
even extends its importance to alternative theories of gravity
(Milgrom \citeyear{milgrom2001shape}; Nipoti et al. \citeyear{nipoti2007vertical}).

One of the first documented explorations of the missing mass problem was conducted
by Oort (\citeyear{oort1932force}), who examined the vertical force on stars in the
vicinity of the Sun. He observed that the vertical force exceeded the expected value
based on visible matter alone. A similar problem was also identified by Zwicky
(\citeyear{zwicky1933rotverschiebung})
through the observation of the mass-to-light ratio in the Coma cluster. Zwicky
found that the mass-to-light ratio derived from dynamical means significantly
exceeded that derived from the luminous matter alone.

The mainstream explanation for the missing mass problem is the existence of non-baryonic
dark matter. This hypothetical matter interacts solely through gravity with
baryonic matter, resulting in an additional gravitational force among the stars.
The adoption of dark matter particles in various astrophysical systems has proven
highly successful in numerous areas. Examples include explaining the rotation
curve of spiral galaxies (Rubin et al. \citeyear{rubin1980rotational}), ensuring
the stability of galactic discs (Ostriker \& Peebles \citeyear{ostriker1973anumerical}),
addressing mass deficiencies in ultrafaint dwarf galaxies (Lin \& Ishak
\citeyear{lin2016ultra}), and contributing to our understanding of cosmological
evolutions (Davis \citeyear{davis2000thecosmological}).

\begin{figure}
    \includegraphics[width=1\linewidth]{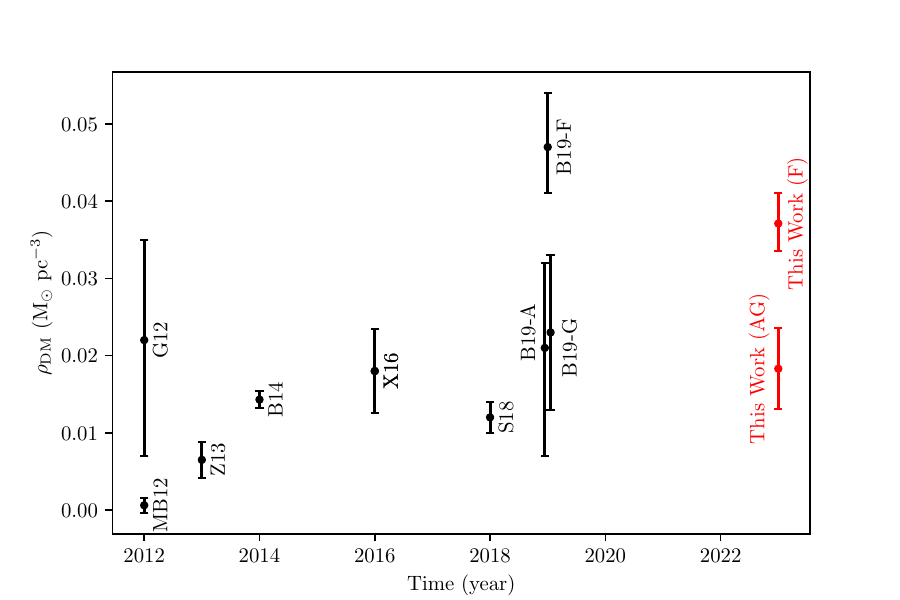}
    \caption{The estimated local dark matter density from several studies in
    the last decade, including our own. The points are from Moni Bidin et al.}
    (\citeyear{bidin2012kinematical}; MB12), Garbari et al.
    (\citeyear{garbari2011limits}; G12), Zhang et al.
    (\citeyear{zhang2013gravitational}; Z13), Bienaym\`e et al.
    (\citeyear{bienayme2014weighing}; B14), Xia et al.
    (\citeyear{xia2016determining}; X16), Sivertsson et al.
    (\citeyear{sivertsson2018localdark}; S18), and Buch et al.
    (\citeyear{buch2019using}; B19) for type A, F, and early G stars.
    \label{fig:local-density}
\end{figure}

The local dark matter density is typically estimated by measuring the vertical
force on stars near the Sun. This is commonly accomplished by measuring the
vertical velocity, a practice that had been studied since Oort (\citeyear{oort1932force}).
Read (\citeyear{read2014local}) provides an extensive review of the methods used to measure
local dark matter. The estimated value of the local dark matter density usually
hovers around $\sim 0.01$ M$_\odot$ pc$^{-3}$, although some recent studies show higher
values, reaching up to $\sim 0.1$ M$_\odot$ pc$^{-3}$ (Bahcall \citeyear{bahcall1984k}).
Fig. \ref{fig:local-density} displays the values of reported local dark matter density.

The local dark matter density is typically estimated by measuring the vertical
force on stars near the Sun. This is commonly accomplished by measuring the
vertical velocity, a practice that had been studied since Oort (\citeyear{oort1932force}).
Read (\citeyear{read2014local}) provides an extensive review of the methods used to measure
local dark matter. The estimated value of the local dark matter density usually
hovers around $\sim 0.01$ M$_\odot$ pc$^{-3}$, although some recent studies
show higher values, reaching up to $\sim 0.05$ M$_\odot$ pc$^{-3}$,
as seen in Fig. \ref{fig:local-density}.

Introduction of dark matter into the universe composition does not come without
its challenges. Among the most significant challenges in the nature of dark matter
are the missing satellite problem and the core-cusp problem. The missing satellite
problem refers to the discrepancy between the number of observed satellite galaxies
around the Milky Way and the number of predicted satellite galaxies in simulations
(Moore et al. \citeyear{moore1999we}). The core-cusp problem refers to discrepancy between density slope 
of the dark matter distribution in the center of galaxies. The dark matter distribution in 
simulations predicts a cusp-like profile (Diemand et al. \citeyear{diemand2005we}), while observations suggest 
a core-like profile (de Blok \citeyear{deblok2010this}). A more recent challenge raised by Haslbauer et al. 
(\citeyear{haslbauer2022the}) shows that the proportion of thin disk galaxies from the cosmological 
$\Lambda$CDM simulation is in significant deficit compared to observations from the Galaxy And Mass 
Assembly (GAMA) survey and Sloan Digital Sky Survey (SDSS). 
These challenges have led to the exploration of
alternative theories to dark matter.

Alternative studies suggest that the apparent effect of missing mass on the
vertical (and radial) force may not be attributed to dark matter itself but rather to
modifications in the gravitational force law. This is typically achieved
by adjusting the Newtonian gravitational force law, introducing slight changes
to the effect of gravity at large distances. One classic example of such a
modification is the \emph{Modified Newtonian Dyanimcs} (MOND) theory, proposed by
Milgrom (\citeyear{milgrom1983modification}). Subsequently, extensive studies have been
conducted to replace dark matter with modified gravity theories, yielding
varying degrees of success (Famaey \& McGaugh \citeyear{famaey2012modified}).

The MOND theory has proven highly successful in elucidating the dynamics of
stars in spiral galaxies (Milgrom \citeyear{milgrom1994dynamics}). This theory mimics the
effects attributed to dark matter, but notably, it does not necessitate the
presence of additional matter. MOND has been successful in explaining the 
flatness of rotation curves of galaxies (Begeman et al. \citeyear{begeman1991extended}) and 
the Baryonic Tully-Fisher relation (McGaugh et al. \citeyear{mcgaugh2000thebaryonic}). Although, MOND
formulation failed to explain the dynamics and lensing of galaxy clusters, such as the infamous 
Bullet Cluster (Clowe et al. \citeyear{clowe2006direct}). The existence of ultra-diffuse galaxies (UDGs) that lack dark matter also supposedly poses a problem for MOND (van Dokkum et al. 
\citeyear{van2018galaxy}), but a thorough analysis including external field effects may solve this problem 
(Kroupa et al. \citeyear{kroupa2018does}; Haghi et al. \citeyear{haghi2019a,haghi2019the}).

There are some differing predictions between MOND theory and dark matter theory. 
Lisanti et al. (\citeyear{lisanti2019testing}) attempt to exploit these differences, particularly in the 
ratio of the vertical force to the radial force in the Milky Way. Their findings reveal that MOND
theory predicts a distinct ratio compared to dark matter theory, providing
a means to test which theory aligns more closely with the observed data.
Another method employed to test the hypotheses of dark matter and MOND involves
comparing the centripetal acceleration of wide binary stars with their
acceleration calculated using classical Newtonian gravity.
Hernandez (\citeyear{hernandez2023internal}) discovered anomalous behavior in the low
acceleration regime, favouring MOND. Chae (\citeyear{chae2023breakdown}) even identified
evidence at a $10\sigma$ level in favor of MOND. Conversely, Pittordis \& Sutherland (\citeyear{pittordis2023wide})
found that standard gravity is more preferred. A more rigorous Bayesian
statistical analysis conducted by Banik et al. (\citeyear{banik2023strong}) revealed results
consistent with Newtonian dynamics, ruling out MOND with a confidence level of
$16\sigma$. However, Hernandez \& Chae (\citeyear{hernandez2024a}) provide a 
thorough critical analysis of the methods used in Banik et al. (\citeyear{banik2023strong}) and 
Pittordis \& Sutherland (\citeyear{pittordis2023wide}), rebutting their claims. 
This complex array of findings indicates the need for further
studies to rigorously test these two competing hypotheses.

In this paper, we will explore the influence of matter distribution (with or whitout
dark matter) on generating gravitational potential and shaping the vertical
distribution of stars near the Sun. We assume that the stars are in dynamical
equilibrium and employ the collisionless Boltzmann equation to model the
distribution of stars (Holmberg \& Flynn \citeyear{holmberg2004local}).
The vertical density profile of tracers and their vertical velocity dispersion are
measured. Subsequently, we attempt to fit the vertical distribution using mass
models that assume isothermal components in dynamical equilibrium. This analysis is
conducted for both dark matter and MOND hypotheses. The comparison of both
best-fittings is then performed by evaluating their Bayes factor.

The paper is organized as follows: In Section \ref{sec:data}, we provide a
detailed description of the data collection and the subsequent data cleaning process.
Section \ref{sec:gravitational-model} outlines the mass model,
theoretical background, and the two hypotheses.
Section \ref{sec:method} explains the methodology employed for data fitting
using Markov Chain Monte Carlo (MCMC). It includes an explanation of the likelihood
function, the prior, and the numerical method used for solving the gravitational
potential and calculating the Bayes factor. Section \ref{sec:results} presents a
the analysis of the results, discussing the dark matter density from the
dark matter hypothesis, the acceleration parameter from the MOND hypothesis, and a
comparison of the two hypotheses using Bayes factors. Finally, Section \ref{sec:conclusion}
summarizes and concludes the findings of our study. Additionally, Section
\ref{app:density-inference} contains the derivation of the likelihood function for
inferring number density.


\section{DATA}\label{sec:data}
The data utilised in this study consist of the three-dimensional positions and
kinematics of stars around the Sun. These data are compiled from Gaia DR3, RAVE 6,
APOGEE-2 (SDSS), GALAH DR3, and LAMOST DR7 catalogues. They are collected either
by downloading the bulk data directly from their repositories or queried using
\texttt{astroquery} through virtual observatory services.

The Gaia catalogue serves as the primary source, providing three-dimensional positions 
($\alpha, \delta, \varpi$) and kinematics ($\mu_\alpha, \mu_\delta, v_r$) of stars. 
Other catalogues are employed to fill in missing radial velocity data from Gaia DR3. 
Table \ref{tab:contribution-distribution} displays the distribution of radial velocity
data from each catalogue. It is evident that the contribution of radial velocity
data from catalogues other than Gaia is not significant, rendering the subsequent
analysis unaffected by data collected solely from Gaia DR3.

\begin{table}
    \caption{Combined contribution of radial velocity data from Gaia DR3, RAVE, APOGEE, GALAH, 
    and LAMOST catalogues.}
    \label{tab:contribution-distribution}
    \begin{center}
        \begin{tblr}{Q[m,c]Q[m,r]}
            \hline
            Catalogue & \SetCell{c}\# \\
            \hline
            GALAH     & 78,055        \\
            APOGEE    & 370,523       \\
            RAVE      & 27,826        \\
            LAMOST    & 442,140       \\
            Gaia DR3  & 23,191,519    \\
            \hline
        \end{tblr}
    \end{center}
\end{table}

Additional information from 2MASS photometry is also utilised to obtain the
colours $J$ and $K_s$ of the stars. The colour $J-K_s$ is used for selecting
the main sequence cut, following a methodology similar to that employed in Bovy's
work (\citeyear{bovy2017stellar}). Infrared photometry is chosen because it is
less susceptible to the effects of interstellar extinction. Hence, for this
study, there is no need to correct for the extinction effect. The 2MASS All-sky
Point Source Catalogue (PSC) encompasses photometry data for 470,992,970 stars
in the near-infrared wavelength ($J$, $H$, and $K_s$ bands).

\subsection{Sample Selection} \label{subsec:procedures}
Firstly, all stars from Gaia DR3 are selected across all directions (all sky).
Subsequently, we crossmatch these stars with the 2MASS PSC to acquire the
colours $J$ and $K_s$. This crossmatching process utilises the 2MASS-Gaia joined
catalogue, a compilation previously conducted by the Gaia team
(Marrese, P. M. et al. \citeyear{marrese2018gaia}) and accessible from the Gaia archive\footnote{\url{https://gea.esac.esa.int/archive/}}.
Both catalogues are outer joined, combining entries whether there are matching
stars from 2MASS to Gaia or not. The resulting catalogue, named Gaia+2MASS,
encompasses astrometry, kinematics, and photometry data for 470,992,970 stars.

The remaining catalogues are obtained by downloading the data from their
respective archives. Fortunately, for each catalogue, there exist a pre-existing
crossmatch with Gaia DR3 conducted by the Gaia team. These crossmatches are
established by simply matching the IDs from the Gaia+2MASS catalogue with the
corresponding join table catalogue. Specifically, there are 450,978 crossmatched
stars from RAVE, 573,733 stars from APOGEE, 588,061 stars from GALAH, and
2,730,472 stars from LAMOST. It is required for the crossmatched catalogue to
include radial velocity data for the stars.

Further filtering is conducted to exclude stars with poor-quality photometry,
astrometry, and outliers. Adopting the criteria outlined in Kim \& L\'epine's work
(\citeyear{kim2022stars}), the following selection criteria are applied to the
stars:
\begin{itemize}
    \item Parallax $\varpi > 0$: Only stars with a positive parallax are
          selected to avoid the complexities associated with distance inference for
          stars with negative parallax.
    \item Parallax error $\sigma_\varpi / \varpi < 0.15$: Stars with good-quality
          parallax data, characterized by a relative error less than 15 per cent, are
          chosen.
    \item $-3.0 < G_{\textup{BP}}-G_{\textup{RP}} < 6.0$ and $3.0 < G < 21.0$:
          This criterion aim to eliminate outliers in the colour-magnitude diagram.
    \item $\sigma \left(F_{\textup{BP}}\right)/F_{\textup{BP}} < 0.10$ and
          $\sigma \left(F_{\textup{RP}}\right)/F_{\textup{RP}} < 0.10$:
          Only stars with good-quality photometry in $G_{\textup{BP}}$ and
          $G_{\textup{RP}}$ bands, characterized by relative errors less than 10 per
          cent, are selected.
    \item \texttt{ruwe} $<1.4$: Only stars with good astrometry, defined by a
          renormalized unit weight error (ruwe) less than 1.4, are chosen.
    \item $1.0 + 0.015\left(G_{\textup{BP}}-G_{\textup{RP}}\right)^2<E<1.3 + 0.06\left(G_{\textup{BP}}-G_{\textup{RP}}\right)^2$:
          This criterion is applied to select stars with a good colour excess $E$.
\end{itemize}
Additional filtering is carried out by incorporating the stars' infrared
photometry from the 2MASS PSC to obtain the highest-quality infrared data.
Following Bovy's methodology (\citeyear{bovy2017stellar}), the criteria for
selecting the best infrared data are:
\begin{itemize}
    \item $2<J<13.5$: Establishing a lower bound to eliminate outliers and an
          upper bound to ensure greater than 99 per cent completeness.
    \item \texttt{ph\_qual} = \texttt{A\_A}: Selecting high-quality photometry
          based on the photometric quality flag.
    \item \texttt{rd\_flg} = 1 or 3: Verifying data quality using the read flag,
          specifically selecting stars with flag values of 1 or 3.
    \item \texttt{use\_src} $ = 1$: Choosing stars that satisfy the criteria for
          good 2MASS data.
\end{itemize}

All the aforementioned criteria are applied to the Gaia+2MASS catalogue,
yielding 29,144,488 stars with Gaia ID, referred to as the pre-combined
catalogue. Additionally, 50,250,399 stars are identified with 2MASS infrared
photometry only, serving as the reference dataset.

In the final step, the pre-combined catalogue is crossmatched with the
remaining catalogues using the Gaia DR3 source ID. This crossmatched catalogue
is denoted as the combined catalogue, encompassing numerous radial velocity data
from various catalogues. The radial velocity data employed in this study
corresponds to the radial velocity with the smallest error from the respective
original catalogue. Stars lacking radial velocity data are excluded from the
combined catalogue, resulting in a dataset comprising 24,110,063 stars with
complete astrometry ($\alpha, \delta, \varpi$), kinematics ($\mu_\alpha,
    \mu_\delta, v_r$), and photometry ($J, K_s$) data.

\subsection{Midplane Selection and Main Sequence Cut}

\begin{figure}
    \centering
    \includegraphics[width=0.8\columnwidth]{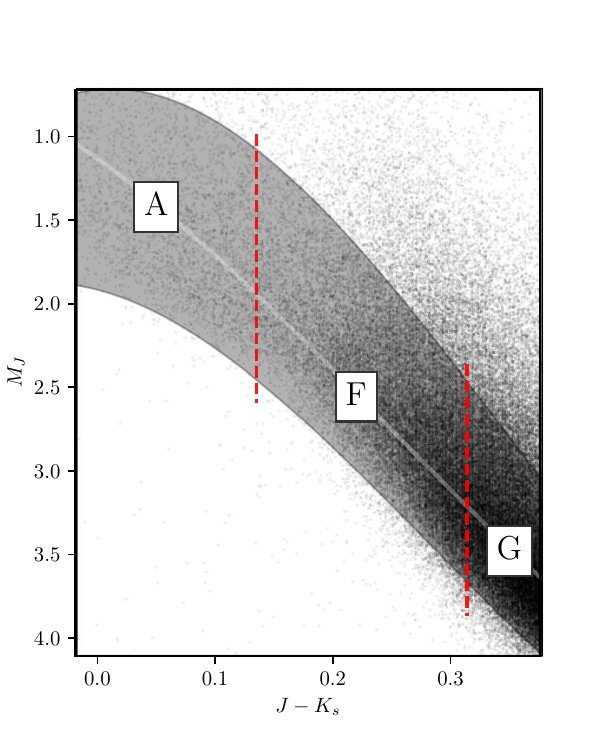}
    \caption{The Hertzsprung-Russell (HR) diagram depicts stars with stellar type A 
    to early-G in the solar neighborhood, with the shaded area denoting the main 
    sequence cut.}
    \label{fig:main-sequence}
\end{figure}

Inferences regarding the stellar dynamics in the Milky Way due to matter
distribution are drawn by analyzing the positions and motions of appropriate
stellar populations, referred to as \textit{tracers}. For this study, we
specifically focus on stars in the solar neighborhood, defining them as stars
within a cylindrical volume in galactic coordinate with a vertical distance 
of $|z| < 200$ pc from the midplane and a radius of $150$ pc. Main
sequence stars are chosen and categorized into multiple groups based on their
colour $J-K_s$. Following the methodology used by Bovy (\citeyear{bovy2017stellar}), we
employ the mean dwarf stellar locus from Pecaut \& Mamajek (\citeyear{pecaut2013intrinsic}) 
and introduce a reasonable thickness to the absolute magnitude $M_J$. Utilising a
similar approach, we define the main sequence by tracing the main branch of the
Hertzsprung-Russell diagram (HRD) of nearby stars, establishing upper and lower
boundaries in absolute magnitude as displayed by the shaded area in Fig.
\ref{fig:main-sequence}.

The stellar populations under consideration in this study must have achieved
dynamical equilibrium within the Galactic disc in recent times. Following the
recommendation of Buch et al. (\citeyear{buch2019using}), we opt for stars 
characterized by lower scale height and a shorter equilibration timescale, 
specifically focusing on younger A stars up to early G stars. Consequently, we 
choose the interval between A0 stars and G3 stars and divide the group into 15 
subgroups, as detailed in Table \ref{tab:stars-group}. The color division for 
each group is carefully selected to ensure a minimum of 1000 stars in the midplane,
guaranteeing robust statistical analyses for each subgroup. The spectral type
of each group is determined from the color $J-K_s$ using the relation provided
by Pecaut \& Mamajek (\citeyear{pecaut2013intrinsic})\footnote{We use version \texttt{2022.04.16},
    \href{https://www.pas.rochester.edu/~emamajek/EEM\_dwarf\_UBVIJHK\_colours_Teff.txt}{https://www.pas.rochester.edu/~emamajek/EEM\_dwarf\_UBVIJHK\_colours\_Teff.txt}}.

\begin{table}
    \caption{Tracer groups of stars utilised in the study, categorized by
        colour. The midplane column indicates the number of stars within
        $|z| < 50$ pc.}
    \label{tab:stars-group}
    \centering
    \begin{tblr}{X[3,c]X[2,c]X[2,c]X[2,c]}
        \hline
        $J-K_s$             & Spectral Type & $|z| < 200$ pc & Midplane \\
        \hline
        $[-0.019$, $0.124]$ & A0 - A8       & 2422           & 1006     \\
        $[0.124$, $0.188]$  & A9 - F1       & 2719           & 1014     \\
        $[0.188$, $0.219]$  & F2 - F4       & 2867           & 1027     \\
        $[0.219$, $0.241]$  & F5            & 2919           & 1014     \\
        $[0.241$, $0.259]$  & F6            & 2970           & 1033     \\
        $[0.259$, $0.275]$  & F7            & 3291           & 1049     \\
        $[0.275$, $0.288]$  & F8            & 3165           & 1032     \\
        $[0.288$, $0.300]$  & F9            & 3315           & 1038     \\
        $[0.300$, $0.312]$  & F9.5          & 3523           & 1060     \\
        $[0.312$, $0.323]$  & F9.5          & 3546           & 1001     \\
        $[0.323$, $0.333]$  & G0            & 3493           & 1007     \\
        $[0.333$, $0.343]$  & G1            & 3580           & 1028     \\
        $[0.343$, $0.353]$  & G1            & 3803           & 1067     \\
        $[0.353$, $0.362]$  & G2            & 3517           & 1044     \\
        $[0.362$, $0.376]$  & G2 - G3       & 3460           & 1529     \\
        \hline
    \end{tblr}
\end{table}

\subsection{Selection Functions}
Stellar information from the combined catalogue is subject to incompleteness
arising from various factors, potentially introducing bias to the inferred
density distribution. To mitigate this issue, it is crucial to correct the
measured density distribution by accounting for the selection function of the
stars. Following the approach outlined by Bovy (\citeyear{bovy2017stellar}), we estimate
the selection function of the stars in the combined catalogue by comparing them
to the assumed complete sample of stars from the reference catalogue (refer to
subsection \ref{subsec:procedures}).

The selection function is defined as the probability of observing a star with
specific properties in the catalogue or survey. Let $S(J, J-K_s)$ represent the
ratio between the number of stars in the combined catalogue and the number of
stars in the reference catalogue with color $J-K_s$ and absolute magnitude $M_J$.
The effective selection function is then defined as:
\begin{align}
    \mathfrak{S}(x,y,z) = \int \rho_{\textup{CMD}}(M_J, c|x,y,z)S(M_J, c)\textup{d}c\textup{d}M_J, \label{eq:effective-selection-function}
\end{align}
where $c=J-K_s$, and $\rho_{\textup{CMD}}(M_J, J-K_s | x, y, z)$ represents the
density distribution of stars in the color-magnitude diagram (CMD) at spatial position
$(x, y, z)$. Equation \eqref{eq:effective-selection-function} can be seen as a
conversion from color-magnitude space completeness to position space completeness.
This equation is then carried to calculate the \emph{effective volume
    completeness}, providing an estimate of the completeness for a given region
$\Pi$, defined as:
\begin{align}
    \Xi(\Pi) = \frac{\int_{\Pi} \mathfrak{S}(x,y,z) \textup{d}^3 x }{\int_{\Pi} \textup{d}^3 x},
\end{align}
where in this case $\Pi$ represents a cylindrical volume with a radius of $150$
pc and varying height $z$ with some thickness $\Delta z$.

For this study, we make the assumption that the density distribution in the
Color-Magnitude Diagram (CMD) is uniform. The values can thus be inferred from
the Hertzsprung-Russell (HR) diagram of stars in the solar neighborhood, as
illustrated in Fig. \ref{fig:main-sequence}.

\begin{figure}
    \includegraphics[width=\columnwidth]{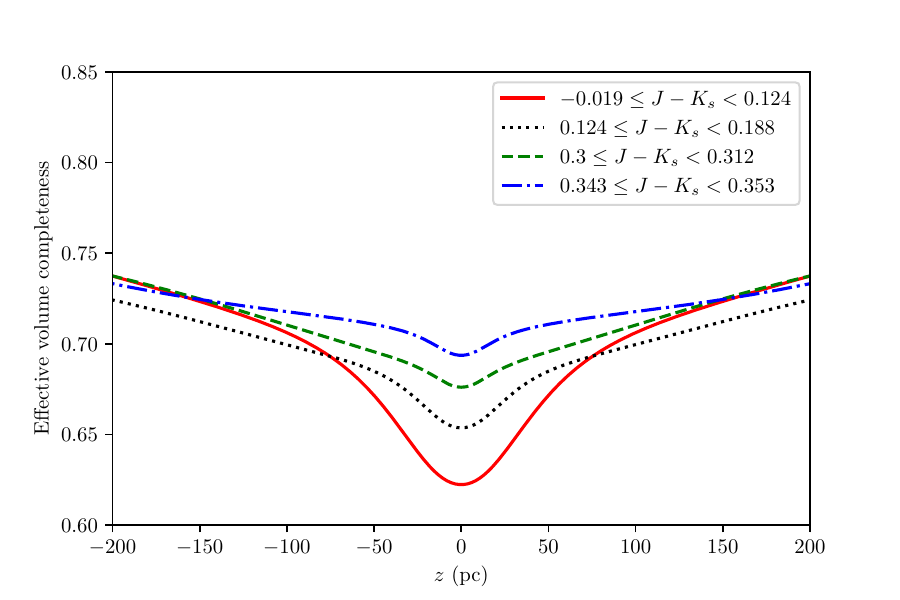}
    \caption{The effective volume completeness function for stars in the solar
        neighborhood. This graph illustrates the behavior for four selected groups
        of stars, with other groups exhibiting similar trends.}
    \label{fig:effective-selection-function}
\end{figure}

The effective volume completeness is employed to adjust the density distribution
of stars inferred from the combined catalogue to the (supposedly) true density
distribution of stars in the solar neighborhood. This correction involves
dividing the density distribution of stars in the combined catalogue by the
effective volume completeness, resulting in what is termed the corrected density
distribution. Figure \ref{fig:effective-selection-function} displays samples
from four groups of stars alongside their corresponding effective volume
completeness. Notably, the effective volume completeness remains relatively
constant for higher values of $|z|$, indicating that correction is essentially
unnecessary for those regions.

\section{MODELS} \label{sec:gravitational-model}
\subsection{Distribution Function}

In a self-gravitating system such as a galaxy, the distribution function (DF)
of a star represents the probability density function of finding a star with a
given position and velocity (or momentum), denoted as $f(\bm x, \bm v)$. In a collisionless
system and under the condition of dynamical equilibrium, the DF adheres to the
collisionless Boltzmann equation (CBE),
\begin{align}
  (\nabla_{\bm{x}}f)\cdot \bm{v} - (\nabla_{\bm{x}}\Phi) \cdot (\nabla_{\bm{v}}f) & = 0 \nonumber                                                    \\
  (\nabla_{\bm{x}}f)\cdot \bm{v}                                                  & = (\nabla_{\bm{x}}\Phi) \cdot (\nabla_{\bm{v}}f), \label{eq:cbe}
\end{align}
where $\Phi$ is the gravitational potential of the system in equilibrium (see Binney \& Tremaine \citeyear{binney2008galactic} p. 277). When
considering the DF of a star in close proximity to the midplane, it can be approximated by
separating the contributions into the vertical and horizontal directions,
\begin{align}
  f(\bm x, \bm v) & = Z(z, w) T(r, v_r, \phi, v_{\phi}), \label{eq:separation-DF}
\end{align}
where $Z(z, w)$ is the DF in the vertical direction, and $T(r, v_r, \phi, v_{\phi})$
is the DF in the horizontal direction. Equation \eqref{eq:cbe} can be expressed
as,
\begin{align}
  w\frac{\partial Z}{\partial z} - \frac{\partial \Phi}{\partial z}\frac{\partial Z}{\partial w} & = 0. \label{eq:cbe-z}
\end{align}
The differential equation \eqref{eq:cbe-z} has a general solution
$Z(z,w)=Z(E_z)\equiv Z(w^2/2\Phi(z))$, where $E_z$ is the energy in the vertical
direction. Thus, the DF in the vertical direction is a function of energy only.
Assuming the star has a vertical velocity dispersion $f_0(w)$ in the midplane,
the DF in the vertical direction can then be expressed as
\begin{align}
  Z(z,w) & = Z(E_z) \nonumber                                                       \\
         & = Z(w^2/2+\Phi(z)) \nonumber                                             \\
         & = Z(w_0^2/2+\Phi(z_0)) \nonumber                                         \\
         & \propto f_0(w_0) \nonumber                                               \\
         & \propto f_0\left(\sqrt{w^2+2[\Phi(z)-\Phi(z_0)]}\right), \label{eq:df-z}
\end{align}
where, $z_0$ and $w_0$ represent the equivalent midplane position and velocity
of the star that has the same energy as the star at $(z,w)$.


If we assume the vertical velocity dispersion $f_0$ follows a Gaussian
distribution, then equation \eqref{eq:df-z} can be solved analytically to obtain
the DF of spatial vertical direction.
\begin{align}
  F_z(z)             & = \int_{-\infty}^{\infty} Z(z, w) \text{d} w\nonumber                 \\
                     & = A \exp\left(-\frac{\Phi(z)}{\sigma_w^2}\right), \label{eq:df-z-sol} \\
  \Rightarrow \nu(z) & = \nu_0 \exp\left(-\frac{\Phi(z)}{\sigma_w^2}\right), \label{eq:nu-z}
\end{align}
where $A$ is a normalization constant, $\sigma_w$ is the vertical velocity
dispersion, and $\Phi(z_0)$ is chosen to be zero. If we only consider $N$
identical stars, then the vertical distribution \eqref{eq:df-z-sol} is
proportional to the expected number density $\nu(z)$ of the stars, leading to
\eqref{eq:nu-z}.

The vertical velocity distribution can be calculated by integrating the DF in
the spatial vertical direction,

\begin{align}
  F_w(w)             & = \int_{-\infty}^{\infty} Z(z, w) \text{d} z\nonumber                          \\
                     & = a^* f_0(w) \nonumber                                                         \\
  \Rightarrow \nu(w) & = a \exp\left[-\frac{\left(w-w_0\right)^2}{2\sigma_w^2}\right]\label{eq:nu-w},
\end{align}
where $a$ is a normalization constant. The vertical velocity distribution
\eqref{eq:nu-w} is a Gaussian distribution with mean $w_0$ and standard
deviation $\sigma_w$.

\subsection{Gravitational Potential}

The gravitational potential of a system can be calculated from the mass density
distribution using Poisson's equation. In a cylindrical volume around the Sun,
Poisson's equation can be expressed as

\begin{align}
  \nabla^2 \Phi                        & = \frac{\partial^2 \Phi}{\partial z^2} + \frac{1}{r}\frac{\partial}{\partial r}\left(r\frac{\partial \Phi}{\partial r}\right) = 4\pi G \rho \nonumber \\
  \frac{\partial^2 \Phi}{\partial z^2} & = 4\pi G\rho_{\textup{eff}}\label{eq:poisson-equation},
\end{align}
where the second term of equation \eqref{eq:poisson-equation} is referred to as
the \textit{rotation curve term} $\mathcal{R}$, which is effectively constant
for a small volume around the Sun. The rotation curve term can be calculated as 
(Binney \& Merrifield \citeyear{binney1998galactic} p. 691)
\begin{align}
  \mathcal{R} & \equiv \frac{1}{4\pi G}\frac{1}{r}\frac{\partial}{\partial}\left(r\frac{\partial \Phi}{\partial r}\right) \nonumber \\
              & \equiv \frac{A^2-B^2}{2\pi G} \label{eq:rotation-curve-term}
\end{align}
where $A$ and $B$ represent the Oort constants. 
The values of $A$ and $B$ are $15.3 \pm 0.4$ km s$^{-1}$ kpc$^{-1}$ and $-11.9 \pm 0.4$ km s$^{-1}$ kpc$^{-1}$,
respectively (Bovy \citeyear{bovy2017galactic}).

For the volume studied in this work, the rotation curve term is approximately
constant and the gravitational potential $\Phi$ is only dependent on $z$.
Therefore, the Poisson's equation \eqref{eq:poisson-equation} can be
written as a second-order ordinary differential equation with respect to $z$,
\begin{align}
  \frac{\textup{d}^2\Phi(z)}{\textup{d}z^2} & = 4\pi G\rho_{\textup{eff}}(\Phi, z), \label{eq:poisson-equation-2nd-order}
\end{align}
where $\rho_{\textup{eff}}(\Phi, z)$ could be a function of $\Phi$ and $z$.
Section \ref{subsec:mass-model} will discuss the mass model used in this work
in detail.

\subsection{Modified Poisson Equation}

In Milgrom's work \citeyear{milgrom1983modification}, he proposed a modification to
Newtonian dynamics by introducing a new constant $a_0$, referred to as the acceleration
parameter. This constant plays a crucial role in scenarios where an object experiences a
low acceleration field. One approach to incorporating the acceleration parameter into
Newtonian dynamics is by ensuring that the circular speed of an object in a low
acceleration regime remains constant:
\begin{align}
  \text{Centripetal acc.} = \frac{V_c^2}{r} \propto \frac{1}{r}.
\end{align}
Such a modification can be achieved by adjusting the gravitating acceleration to follow
\begin{align}
  g = \sqrt{g_N a_0},
\end{align}
where $g_N$ represents the classical Newtonian gravitational acceleration.

Concerning this phenomenology, MOND introduces a different form of Poisson's equation,
as formulated by Bekenstein \& Milgrom (\citeyear{bekenstein1984doesthemissing}) for
AQUAL theory:
\begin{align}
  4\pi G\rho = \nabla \cdot \left[\mu\left(\frac{|\nabla \Phi|}{a_0}\right)\Phi\right], \label{eq:poisson-modified}
\end{align}
where $a_0$ represents the MOND acceleration parameter constant, and $\mu$ is the
MOND $\mu$-interpolating function. The $\mu$-interpolating function is defined as
\begin{align}
  \mu(x) & = \begin{cases}
               1 & \text{if } x \gg 1 \\
               x & \text{if } x \ll 1
             \end{cases}. \label{eq:interpolating-function}
\end{align}
Equation \eqref{eq:interpolating-function} ensures that the modified Poisson's
equation reduces to the Newtonian Poisson's equation when the acceleration is
much larger than the MOND acceleration parameter $a_0$. In small volumes, where
the total acceleration $|\nabla \Phi |$ is approximately constant, the modified
Poisson's equation \eqref{eq:poisson-modified} can be simplified to:
\begin{align}
  \nabla \cdot \left[\mu\left(\frac{|\nabla \Phi|}{a_0}\right)\Phi\right]  &= 4\pi G\rho \nonumber \\
  \mu\left(\frac{|\nabla \Phi|}{a_0}\right) \nabla ^2 \Phi & = 4\pi G\rho \nonumber \\
  \nabla ^2 \Phi &= \frac{4\pi G \rho}{\mu_0}, \label{eq:poisson-modified-simplified}
\end{align}
where $\mu_0=\mu\left(|\nabla \Phi|/a_0\right)$ is the value of the interpolating function at the Sun's position.

Both Poisson equations \eqref{eq:poisson-equation-2nd-order} and
\eqref{eq:poisson-modified-simplified} are similar, with a minor difference
represented by a factor of $\mu_0$. This value can be interpreted as the
additional boost of the gravitational potential due to the MOND effect. In the
dark matter hypothesis, this additional boost is generated by additional mass
density from nonluminous dark matter. Essentially, the MOND effect can be
mimicked by adding extra mass density to the system. The two equations can
result in the exact same gravitational potential if the additional missing mass
follows the baryonic mass distribution. Fortunately, most dark matter models
predict that the dark matter distribution does not precisely follow the baryonic
mass distribution (Navarro et al. \citeyear{navarro1997a}). As a result, the two
equations should yield different
gravitational potentials in theory. The extent of the difference between the two
equations is determined by the value of $\mu_0$. If $\mu_0$ is close to 1, it
becomes extremely challenging to distinguish between the hypotheses of dark
matter and MOND.

The exact function of the $\mu$-interpolating function is not precisely known.
However, for many studies focusing on the low acceleration regime, the specific
form of the interpolating function is often not critical. In this study, where
the acceleration is not low enough to be in the low acceleration regime, the
exact function of the interpolating function becomes crucial for inferring the
value of the acceleration parameter $a_0$. We consider two common interpolating
functions: the simple and the standard (Famaey \& McGaugh \citeyear{famaey2012modified}). 
There are other interpolating functions, such as the RAR (McGaugh et al. \citeyear{mcgaugh2016radial})
that is currently being considered in MOND literature, but only has a well-defined form in the $\nu$-interpolating function. 
Equation \eqref{eq:interpolating-function-formulas} provides the formulas for the relevant interpolating functions under 
consideration.
{Green}\begin{align}
  \mu(x) & = \begin{cases}
             \displaystyle \frac{x}{1+x} & \text{Simple} \\
             \vspace{0.01em} \\
             \displaystyle \frac{x}{\sqrt{1+x^2}} & \text{Standard}
           \end{cases}. \label{eq:interpolating-function-formulas}
\end{align}
\subsection{Mass Model} \label{subsec:mass-model}

The model employed in this study encompasses analytical functions for the
vertical distribution of stars \eqref{eq:nu-z} and the vertical velocity
distribution of stars \eqref{eq:nu-w}. This model is derived by solving the
Poisson's equation (\ref{eq:poisson-equation} or
\ref{eq:poisson-modified-simplified}) with the provided mass density
distribution. The mass density distribution for this study is the sum of $N_{\textrm{b}}$
baryon components and a constant dark matter (DM) halo contribution in the midplane.
The baryon mass density $\rho_{\textrm{b}}$ is described by the Bahcall model, which
includes isothermal gas, stars, and stellar remnants (Bahcall \citeyear{bahcall1984the,bahcall1984self,bahcall1984k}).

\begin{align}
  \rho_{\textrm{b}}(z) & = \sum_{i=1}^{N_b} \rho_i(0)\exp \left(-\frac{\Phi(z)}{\sigma_{z,i}^2}\right), \label{eq:bahcall}
\end{align}
where $\rho_i(0)$ represents the midplane mass density of the $i$-th component at
the midplane, and $\sigma_{z,i}$ denotes the vertical velocity dispersion of the
$i$-th component. Table \ref{tab:bahcall-component} provides the parameters of
the Bahcall model utilized in this study.

Because the region of interest is in the vicinity of the Sun, the exact mass
distribution of the Milky Way globally is not crucial. Especially the effect of
galactic bulge is assumed to be constant in the region of interest. This effect
is included in the absolute value of the galactic potential $\Phi$. The external field
effect we exclude is the one originating outside the Milky Way. This is beyond the scope of this study.

\begin{table}
  \caption{Baryonic mass components employed in this study. The table is
    extracted from table 2 of Buch et al. (\citeyear{buch2019using}).
  }
  \label{tab:bahcall-component}
  \begin{center}
    \begin{tblr}{Q[m,c]Q[m,c]Q[m,c]}
      \hline
      Component                                      & $\rho_{i,0} \left[\textrm{M}_\odot/\textup{pc}^2\right]$ & $\sigma_{z,i} \left[\textup{km/s}\right]$ \\
      \hline
      Molecular gas (H$_2$)                          & $0.0104 \pm 0.00312$                                     & $3.7 \pm 0.2$                             \\
      Cold atomic gas (H$\,\text{\scriptsize I}(1)$) & $0.0277 \pm 0.00554$                                     & $7.1 \pm 0.5$                             \\
      Warm atomic gas (H$\,\text{\scriptsize I}(2)$) & $0.0073 \pm 0.00070$                                     & $22.1 \pm 2.4$                            \\
      Hot ionized gas (H$\,\text{\scriptsize II}$)   & $0.0005 \pm 0.00003$                                     & $39.0 \pm 4.0$                            \\
      Giant stars                                    & $0.0006 \pm 0.00006$                                     & $15.5 \pm 1.6$                            \\
      $M_V < 3$                                      & $0.0018 \pm 0.00018$                                     & $7.5 \pm 2.0$                             \\
      $3< M_V < 4$                                   & $0.0018 \pm 0.00018$                                     & $12.0 \pm 2.4$                            \\
      $4< M_V < 5$                                   & $0.0029 \pm 0.00029$                                     & $18.0 \pm 1.8$                            \\
      $5< M_V < 8$                                   & $0.0072 \pm 0.00072$                                     & $18.5 \pm 1.9$                            \\
      $M_V > 8$ (M dwarfs)                           & $0.0216 \pm 0.00280$                                     & $18.5 \pm 4.0$                            \\
      White dwarfs                                   & $0.0056 \pm 0.00100$                                     & $20.0 \pm 5.0$                            \\
      Brown dwarfs                                   & $0.0015 \pm 0.00050$                                     & $20.0 \pm 5.0$                            \\
      \hline
    \end{tblr}
  \end{center}
\end{table}

The DM halo contribution is assumed to be constant in the vicinity of the Sun,
as shown by Garbari et al. (\citeyear{garbari2011limits}) for small volume around 
the Sun up to $\sim$ 750 pc.
So the total mass density distribution is given by
\begin{align}
  \rho(z) & = \rho_{\textrm{b}}(z) + \rho_{\textrm{\scriptsize DM}}. \label{eq:total-density}
\end{align}
Based on the mass density distribution, the gravitational potential is
calculated by solving the Poisson equation \eqref{eq:poisson-equation} or
\eqref{eq:poisson-modified-simplified}. This potential is then used to compute
the vertical distribution of stars \eqref{eq:nu-z} and the vertical velocity
distribution of stars \eqref{eq:nu-w}.

Solving the Poisson equation with this mass model is not trivial because
it is not analytically solvable or separable. Numerical methods are needed to
solve the Poisson equation. Fundamentally, from the Poisson equation, given
mass model $\rho$, there is a unique gravitational potential $\Phi$.

In this study, we explore three hypotheses: the baseline hypothesis involves
Classic Newtonian gravity without Dark Matter (NO), while the other two hypotheses
consider Newtonian gravity with dark matter (DM) and Modified Newtonian Dynamics
(MOND). The DM model incorporates the DM halo contribution into the mass density
distribution, while the MOND model relies solely on baryonic mass contributions.
Despite the exclusion of the DM halo contribution in the MOND model's equation
\eqref{eq:total-density}, it introduces the MOND effect into the Poisson equation
\eqref{eq:poisson-modified-simplified}. Equation \eqref{eq:poisson-modified-simplified}
can be perceived as the traditional Poisson equation with baryonic mass density,
scaled by a factor of $1/\mu_0$ times the baryonic matter. Essentially, for our
specific study, the MOND model is equivalent to a scenario without a DM model
using classical Newtonian gravity, albeit with adjusted mass density.

\section{METHODS} \label{sec:method}

\subsection{Density Inference}
The most straightforward method for inferring the density distribution of stars
from their spatial distribution is to count the number of stars within a
specified volume. While this approach is rudimentary, its simplicity is a
powerful tool, particularly when dealing with a large number of stars. However,
a drawback of this method is the omission of uncertainties in star positions,
which may be significant in certain cases.

Previous attempts, as demonstrated by Buch et al. (\citeyear{buch2019using});
Xia et al. (\citeyear{xia2016determining}), Sivertsson (\citeyear{sivertsson2018localdark}),
among others, have aimed to infer density uncertainty
by employing Poisson error or simple standard deviation within a given volume.
While this approach is superior to having no uncertainty estimation, it often
assumes that the distribution follows a Gaussian (or even log-Gaussian) shape,
which is not always accurate. For a large number of stars, the Poisson
distribution can be well approximated by a Gaussian distribution. However,
when dealing with a small number of stars, assuming a
Gaussian distribution might lead to inaccuracies. In such cases, the distribution should be
considered back to Poisson distribution.

While the Poisson error in a given volume accounts for uncertainty, it does not
consider the individual star's position uncertainty. For datasets with high-quality
parallax data, individual star position uncertainties may be negligible. However,
in our study, we aim to develop a method applicable to any type of data, including
those with poor quality and significant uncertainty. Appendix \ref{app:density-inference}
outlines the methodology for inferring the density distribution of objects from their
'position'. Using this approach, we can infer the density distribution of stars from
their spatial distribution, taking into account the uncertainty in their positions.
In other words, for a given distance $z$ from the midplane, there is a probability
density function $P(\nu,z)$ representing the likelihood of the density being $\nu$.

We employ the technique proposed by Bailer-Jones (\citeyear{bailer-jones2015estimating})
to obtain the probability distribution function of the distance $r$ with a constant volume
density prior:
\begin{align}
    p(r|\varpi, \sigma_{\varpi}) \propto r^2\exp\left[-\frac{1}{2\sigma_{\varpi}^2}\left(\varpi - \frac{1}{r}\right)^2\right]. \label{eq:pdf-r}
\end{align}
Subsequently, we convert this distribution to the vertical distance $z$,
\begin{align}
    p(z|\varpi, \sigma_{\varpi}, b) \propto z^2\exp\left[-\frac{1}{2\sigma_{\varpi}^2}\left(\varpi - \frac{\sin b}{z}\right)^2\right], \label{eq:pdf-z}
\end{align}
where $(\varpi, \sigma_{\varpi})$ represent the parallax and its error,
$z$ denotes the vertical distance from the Galactic midplane, and $b$ is the
Galactic latitude.

A similar approach is used to determine the probability distribution
function of vertical velocity. The formula used to calculate $w$ from its
proper motion and radial velocity is given by:
\begin{align}
    w(\mu_b, b, v_r, r) = \kappa \mu_b r \cos b + v_r \sin b,
\end{align}
where $\kappa=4.74$ km s$^{-1}$ pc$^{-1}$ arcsec yr$^{-1}$, $\mu_b$ is the
proper motion in the galactic latitude direction, and $v_r$ is the radial
velocity.

We generate 100,000 samples of ($\varpi, \mu_b, v_r$) for each star, then calculate
their $z$ and $w$ values. Subsequently, we bin both $z$ and $w$ for all star groups
and calculate their weights $W$ according to the methods outlined in Appendix
\ref{app:density-inference}. Fig \ref{fig:fit} shows one such inference for the
tracer of the first group (A0-A8).

\subsection{Likelihood Function}

The likelihood function is divided into two parts: the likelihood function
of the vertical number density $z$ and the likelihood function of the vertical
velocity dispersion $w$. Given a set of parameters $\zeta$ that describe the
vertical distribution of stars, the likelihood function of finding $N(z_i|\zeta)$
stars in the $i$-th bin is expressed as
\begin{align}
    \mathcal{L}(N(z_i|\zeta)) & = \sum_{n=0|z=z_i}^{\infty}W(n)\textup{Poisson}\left(n|N(z_i|\zeta)\right), \label{eq:likelihood-z}
\end{align}
where $W(n)$ represents the weight of finding $n$ stars in the $i$-th bin,
calculated from the data collected in Section \ref{sec:data}. The value of
$N(z_i|\zeta)$ is calculated using \eqref{eq:nu-z}. The term $\textup{Poisson}(n|N)$
is the probability density function of the Poisson distribution, representing
the likelihood of finding $n$ stars given the expected number of stars $N$.
For a more detailed derivation, refer to Appendix \ref{app:density-inference}.

The same function is used to calculate the likelihood function of finding
$N(w_i|\xi)$ stars in the $i$-th bin of the velocity distribution, where
$\xi$ is a set of parameters defining the vertical velocity distribution of
the stars. Therefore,

\begin{align}
    \mathcal{L}(N(w_i|\xi)) & = \sum_{n=0|z=z_i}^{\infty}W(n)\textup{Poisson}\left(n|N(w_i|\xi)\right). \label{eq:likelihood-w}
\end{align}

The total likelihood function is the product of the likelihood function for
the vertical number density and the likelihood function for the vertical
velocity dispersion. The complete set of free parameters $\theta$ encompasses
both the combined dynamics parameters $\zeta$ and the kinematics parameters $\xi$.

\subsection{Priors}

The prior distribution of the parameters $\zeta$ and $\xi$ is summarized in
Table \ref{tab:prior-dynamics} and Table \ref{tab:prior-kinematics},
respectively. We assume that for each group, there are actually two populations
of stars: one with a broader velocity dispersion and another with a narrower
velocity dispersion. Most likely, these two populations correspond to the thin
disc and the thick disc and/or halo. The thin disc is assumed to be the
dominant population, so its contribution is assumed to be higher than that
of the contaminant component. This assumption is based on the fact that the
thin disc is the dominant population in the solar neighborhood. Adopting
this approach, we observe that the fitting curve is improved compared to using
only one population.

Only two types of prior distributions are used: uniform distribution and normal
distribution. In Table \ref{tab:prior-dynamics} and Table
\ref{tab:prior-kinematics}, the ``Constraints'' column is defined as lower and
upper bounds $[a, b]$ if the distribution is uniform, and mean and standard
deviation $[\mu, \sigma]$ if the distribution is normal.

We also impose restrictions on the joint prior distribution of the parameters
$\sigma_{w,1}$ and $\sigma_{w,2}$ to be $\sigma_{w,1} < \sigma_{w,2}$ and on
the parameters $a_1$ and $a_2$ to be $a_1 > a_2$. Parameter $a$ is the sum of
$a_1$ and $a_2$. These restrictions are implemented based on the assumption
that the thin disc is the dominant population and that the thin disc has a
narrower velocity dispersion than the contaminant population.

\begin{table}
    \caption{Summary of kinematic priors, with units indicated in parentheses
        corresponding to the argument of logarithm.}
    \label{tab:prior-kinematics}
    \centering
    \begin{tblr}{X[1,c]X[1,c]X[3,r]X[1,l]}
        \hline
        Parameter           & Distribution & Contraints                                               & Unit   \\
        \hline
        $w_0$               & Uniform      & $[-15, 0]$                                               & km/s   \\
        $\log \sigma_{w,i}$ & Uniform      & $[\ln 3, \ln 50]$                                        & [km/s] \\
        $\log a$            & Normal       & $[\ln \textup{max}(N(w_i)), 2]$                          & -      \\
        $\log a_2$          & Uniform      & $[\ln \textup{max}(N(w_i))-5, \ln \textup{max}(N(w_i))]$ & -      \\
        \hline
    \end{tblr}
\end{table}

\begin{table}
    \caption{Summary of dynamic priors. The constraints for baryon parameters
        are referenced to Table \ref{tab:bahcall-component}.}
    \label{tab:prior-dynamics}
    \centering
    \begin{tblr}{X[1,c]X[1,c]X[3,r]X[1,l]}
        \hline
        Parameter                  & Distribution & Contraints                                       & Unit                  \\
        \hline
        $\rho_{\textrm{b},i}$      & Normal       & Table \ref{tab:bahcall-component}, 12 components & M$_{\odot}$ pc$^{-3}$ \\
        $\sigma_{z,i}$             & Normal       & Table \ref{tab:bahcall-component}, 12 components & km s$^{-1}$           \\
        $\rho_{\textup{\tiny DM}}$ & Uniform      & [-0.05, 0.15]                                    & M$_{\odot}$ pc$^{-3}$ \\
        $\ln \mu_0$                & Uniform      & [$\ln 0.1, \ln 2.0$]                             & -                     \\
        $\ln \nu_0$                & Normal       & [$\ln \textup{max}(\vec N(z_i))$, 3]             & -                     \\
        $\mathcal{R}$              & Normal       & [3.4, 0.6]$\times 10^{-3}$                       & M$_{\odot}$ pc$^{-3}$ \\
        $z_{0}$                    & Uniform      & [-150, 150]                                      & pc                    \\
        \hline
    \end{tblr}
\end{table}

\subsection{MCMC}

We make use of the Markov Chain Monte Carlo (MCMC) method to sample the posterior
distribution of the parameters $\theta$. The MCMC simulations are
conducted using \texttt{hammer-and-sample}\footnote{\href{https://crates.io/crates/hammer-and-sample}{https://crates.io/crates/hammer-and-sample}}
by Adam Reichold, a Rust implementation of the \emph{affine invariant ensemble sampler}
introduced by Goodman \& Weare (\citeyear{goodman2010ensemble}), which is a reimplementation of
\texttt{emcee} Python library by Foreman-Mackey et al. (\citeyear{foreman2013emcee}).

For each group, the MCMC is run for 50,000 steps with 10 $\times$ \texttt{ndim}
walkers, where \texttt{ndim} is the total number of dimension of $\theta$ parameters for each model,
$=33$ for DM or MOND model, and \texttt{ndim} $=32$
for classic NO model (see Table \ref{tab:prior-kinematics} and \ref{tab:prior-dynamics}). The first 4,000 iterations are discarded as burn-in, and the
samples are thinned by a factor of 20.

The MCMC program is implemented in the \texttt{Rust} programming language to
enhance performance and reliability. With 33 free parameters for each group
in the DM model, spanning 15 groups based on infrared color, and considering 2
gravitational models, a substantial amount of computation is involved. Although
the MCMC program is easy to implement, optimizing it can be quite challenging.
\texttt{Rust} demonstrates superior performance, outperforming even well-written
\texttt{Python} code. While \texttt{Rust} may not be as user-friendly as
\texttt{Python}, and its ecosystem is not as mature in Science and Engineering,
the significant performance gains justify the effort. Despite this, the program
is also ported to \texttt{Python} for the sake of easier data analysis and
visualization.

\subsection{Solving the Poisson Equation}

The solution to the Poisson equation is achieved using the Dormand-Prince
method (Dorman \& Prince, \citeyear{dormand1980afamily}) of order 5(4) with dense
output of order 4 (\texttt{Dopri5}). This method is implemented in the \texttt{Rust}
crate \texttt{ode\_solvers}\footnote{\href{https://crates.io/crates/ode\_solvers}{https://crates.io/crates/ode\_solvers}}
by Sylvain Renevey. The choice of this method is driven by its speed, ease of
use, and direct integration into \texttt{Rust}. The output from the integration
seamlessly interfaces with the rest of the program in \texttt{Rust}, eliminating
the need for data structure conversion or cross-language interfaces. The native
code compilation enhances the program's performance blazingly fast.

The general idea of the method involves providing the initial conditions
$\Phi(0)=0$ and $\partial_z \Phi(0) = 0$, along with a set of parameters
$\theta$. The method then performs the integration of the Poisson equation at discrete
values of $z$, yielding the values of $\Phi(z)$ and $\partial_z \Phi(z)$ at
those points. The resulting integration is subsequently interpolated using a
simple linear interpolation method to obtain the value of $\Phi(z)$ at any
arbitrary $z$. This calculated $\Phi(z)$ value is then used to determine
the vertical number density of stars using \eqref{eq:nu-z}.

\subsection{Bayes Factor Estimation}

The computation of the Bayes factor from the MCMC chain involves employing the
approach presented in Robert \& Wraith (\citeyear{robert2009computational}). The
\emph{harmonic means}
method is utilized to estimate the evidence term for each MCMC simulation,
given by
\begin{align}
    E \approx N \left(\sum_{i=1}^N \frac{\phi(\theta_i)}{\mathcal{L}(\theta_i)p_0(\theta_i)}\right)^{-1},
\end{align}
where $N$ denotes the number of samples obtained from the MCMC simulations,
and $\phi(\theta_i)$ is any arbitrary probability distribution function over
$\theta$. While a simple harmonic means method estimates the evidence term by
selecting $\phi = p_0$, it can be risky due to the instability of the
approximation (Newton \& Raftery \citeyear{newton1994approximate}). In our case, we implement the
uniform distribution $\phi(\theta)$ with boundaries around the peak of the
posterior distribution $p(\theta)$, following the strategy outlined in
Robert \& Wraith (\citeyear{robert2009computational}), which stabilizes the approximation. It is
crucial to note that for this method to be effective, the proper prior
distribution function $p_0$ must be used. The Bayes factor is defined as the
ratio between the evidence terms of different models, $B_{12} = E_1/E_2$.
We calculate the approximation 10 times, then take the mean and the standard
deviation as the estimated value.


\section{RESULTS AND DISCUSSION} \label{sec:results}
After running the MCMC for each group and model, the results are in the form
of a 3-dimensional array with respect to the number of walkers, number of parameters,
and number of steps. These results are then flattened (combining all walkers)
to obtain the full samples of the posterior distribution functions $\theta$ (both kinematics and dynamics parameters).
The dark matter density $\rho_{\textrm{\tiny DM}}$ and boost interpolating
function value $\mu_0$ are extracted by constructing their histograms.
Appropriate functions are used to obtain their central values and credible intervals.

\subsection{Dark Matter Density} \label{subsec:dark-matter-density}

\begin{figure*}
    \includegraphics[width=\textwidth]{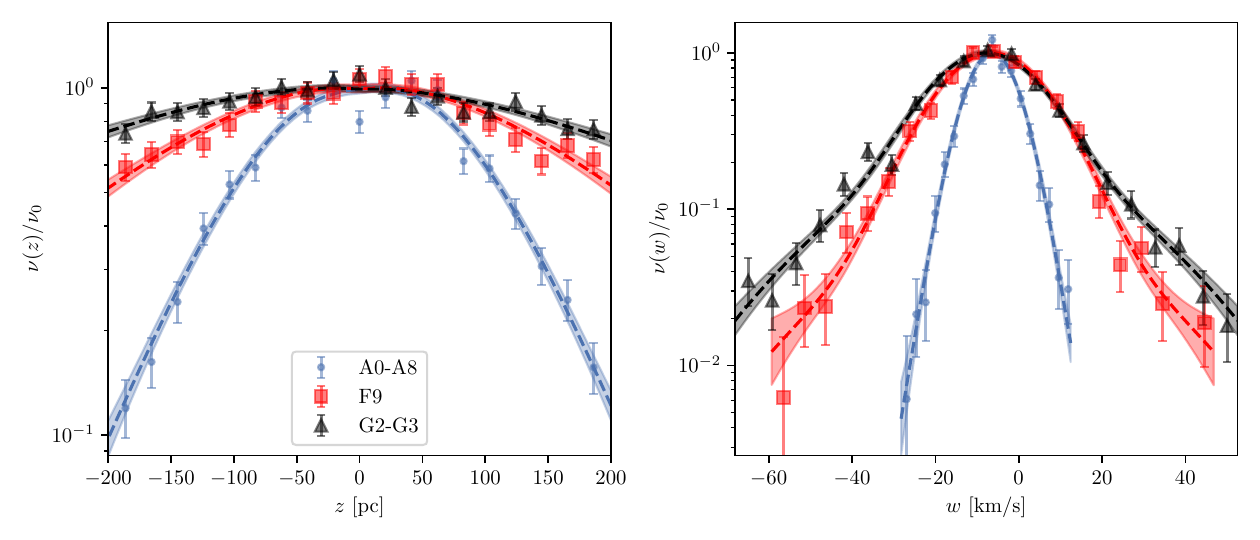}
    \caption{Density inference (error-bars) and best-fitting curve (shaded area) are shown for the
        vertical profile density (left) and vertical velocity dispersion (right)
        for three representative tracer groups (A0-A8, F9, G2-G3). All uncertainties are presented in 68.}
    \label{fig:fit}
\end{figure*}

\begin{figure}
    \includegraphics[width=\columnwidth]{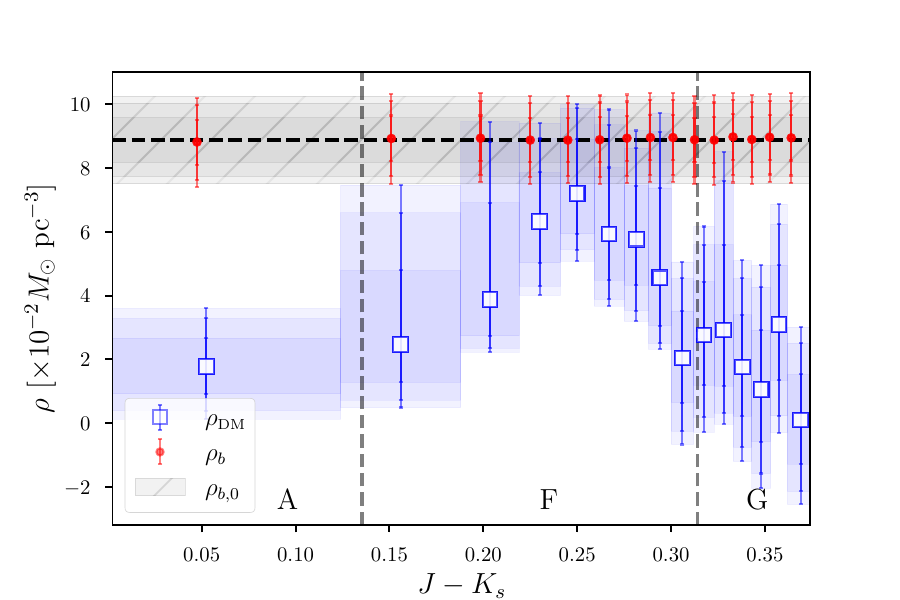}
    \caption{Inferred dark matter density for each group of stars.
        The vertical dashed line separates different spectral types,
        while the horizontal dashed line represents the initial total
        baryonic matter density from Table \ref{tab:bahcall-component},
        with the gray diagonal shaded area at the top indicating the uncertainty. 
        The red dotted error bar plot represents the inferred baryonic matter density, 
        and the blue squared error bar plot depicts the inferred dark matter density from
        the fitting process. All uncertainties are presented in 68, 90, and 95
        per cent credible intervals.}
    \label{fig:dark-matter}
\end{figure}

The full set of estimated parameters $\theta$ is best-fitted to the inferred density
profile (vertical number density and vertical velocity dispersion) using MCMC 
(Fig. \ref{fig:fit}). We can see that the vertical profile of early type tracer is 
more narrowed than the late type and so does the vertical dispersion. These fitting procedure
reveal that the contribution of dark matter (parameter $\rho_{\textrm{\tiny DM}}$ in $\theta$) in the 
halo is approximately on the order of $\sim 0.01 \, \textrm{M}_{\odot} , \textrm{pc}^{-3}$ (Fig. \ref{fig:dark-matter}). 
The results exhibit fluctuations and are not single-valued, as illustrated in Fig. \ref{fig:dark-matter}.
The figure also displays the inferred total baryonic matter density profile, which aligns
with the initial values provided in Table \ref{tab:bahcall-component}. This
indicates that the fitting process accurately captures the contribution of
the known baryonic matter.

In broad terms, the inferred dark matter densities from A-type and G-type stars
exhibit agreement, hovering around $\sim 0.01-0.03$ M$_{\odot}$ pc$^{-3}$,
while the value for F-type stars is notably higher, reaching up to $\sim 0.07$
M$_{\odot}$ pc$^{-3}$. The combined dark matter density for all groups is
$(3.71^{+0.39}_{-0.36})\times 10^{-2}$ M$_{\odot}$ pc$^{-3}$, while if F-type stars
are excluded, the combined infered dark matter density becomes $(1.83^{+0.53}_{-0.52})\times 10^{-2}$
M$_{\odot}$ pc$^{-3}$. Theoretically, the dark matter density should be
consistent across all spectral types, as the dark matter gravitational potential
contribution is expected to be independent of the spectral type of stars. The
implication of this result could stem from either the unsuitability of the dark
matter model employed in this study or the violation of some assumptions made
in the analysis. Let us discuss the possible causes of this result.

Firstly, it's important to note that the dark matter model employed in this
study is relatively simple, assuming a constant density up to 200 pc above
and below the galactic plane. However, simulations of Milky Way-like galaxies,
as indicated by Garbari et al. (\citeyear{garbari2011limits}), suggest that the
dark matter density remains approximately constant up to 1 kpc above and below
the galactic plane, even after 4 Gyrs of evolution. While the simulation
may not perfectly reflect the real Galaxy, it provides valuable insights
into the nature of dark matter distribution. Consequently, it seems
unlikely that refining the dark matter density model would significantly alter
the obtained results.

Secondly, among the various assumptions made in this study, two stand out as
potential contributors to the observed peculiar results. The first assumption
involves the baryon mass distribution, following the Bahcall model, which
consists of several isothermal components. The isothermality assumption implies
that the velocity dispersion remains constant up to 200 pc. Fortunately, simulations
of Milky Way-like galaxies, as suggested by Garbari et al. (\citeyear{garbari2011limits}), indicate
that the velocity dispersion is approximately constant up to 1 kpc above and
below the galactic plane. Thus, this assumption is likely valid. The second
assumption that raises suspicion is the assumption of dynamical equilibrium,
implying that the stars under consideration maintain the same distribution
function after years of interaction and mixing. Achieving dynamical equilibrium
requires that the stars are old enough to interact and mix but not so old that
they have been evaporated from the galaxy. Furthermore, dynamical equilibrium
can only be sustained if there are no significant perturbations to the system,
such as a merger with other galaxies (Malhan et al. \citeyear{malhan2022global}). This might be the
primary cause of the fluctuating results in the inferred dark matter density for
different groups of stars within distinct color ranges.

The lack of convergence in the results for dark matter density strongly
suggests that the sampled data is not in dynamical equilibrium. This
finding aligns with the results reported by Buch et al. (\citeyear{buch2019using}), indicating
that the dark matter density differs for various spectral types of stars.
To address this issue, a modification to the collisionless Boltzmann
equation may be necessary to account for the effects of disequilibrium.
One potential approach involves introducing external wave-like perturbations
to the distribution function, as proposed by (Banik et al. \citeyear{banik2016galactoseismology}).
This presents an intriguing avenue for future research.

\subsection{Acceleration Parameters} \label{subsec:acceleration-parameters}

Fitting the data without including dark matter contribution reveals that the
inferred baryonic matter density consistently exceeds the initial value from
Table \ref{tab:bahcall-component}. This suggests the presence of another matter
component, such as dark matter, or the possibility that the additional matter
is an effect of modified gravity.

Equation \eqref{eq:poisson-modified-simplified} shows the relationship
between baryonic matter and the additional $\mu_0$ boost from MOND theory.
Our analysis yields a value of $\mu_0$ around $\sim 0.64 \pm0.03$ for the
Sun's location in the galaxy, without delving into the specifics of MOND theory.

A similar method to test MOND theory was conducted by Lisanti et al. (\citeyear{lisanti2019testing}),
revealing that the value of the $\nu$-interpolating function is
$1.44^{+0.13}_{-0.08}$. Our equivalent $\nu$-interpolating function value is
$1.56\pm 0.06$, which falls within the uncertainty range. This suggests
that our analysis results are consistent with their findings.

\begin{figure}
    \includegraphics[width=\columnwidth]{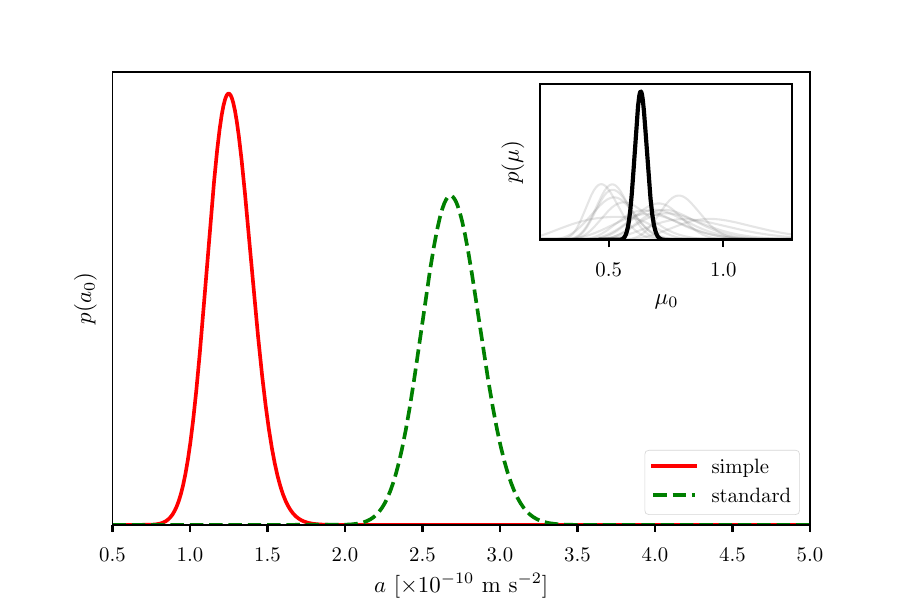}
    \caption{The acceleration parameter probability density derived from
        the infered $\mu_0$. The curves in inset plot are the infered $\mu_0$
        for each groups and dark bold curve is the combined values for all groups,
        around $0.64$.
    }
    \label{fig:mu0}
\end{figure}

MOND theory places particular emphasis on determining the value of the
acceleration parameter $a_0$. Converting the value of $\mu_0$ into acceleration
parameters $a_0$ requires a specific interpolating function. We use two
interpolating functions: the simple one and the standard one, as outlined in
equation \eqref{eq:interpolating-function-formulas}. The resulting value of $a_0$ for the
simple interpolating function is approximately $(1.26^{+0.14}_{-0.13})\times 10^{-10}$ m s$^{-2}$,
while for the standard interpolating function, it is around $(2.69^{+0.19}_{-0.17})
    \times 10^{-10}$m s$^{-2}$. Fig. \ref{fig:mu0} shows the probability
density function for the simple and standard interpolating functions.
An intriguing observation is that the values of $a_0$ for these
two interpolating functions differ significantly. Upon closer inspection,
it is noted that the uncertainties of both results do not overlap, suggesting
the potential to distinguish between the two models.

The inferred acceleration parameter $a_0$ from galactic rotation curve analysis, as found by
Gentile et al. (\citeyear{gentile2011things}) has a value of $a_0$ to be $(1.27\pm 0.30) \times 10^{-10}$ m s$^{-2}$ for the standard 
interpolating function and $(1.22\pm 0.33) \times 10^{-10}$ m s$^{-2}$ for the simple interpolating function. Begeman et al.
(\citeyear{begeman1991extended}) found the average value of $a_0$ using standard interpolating function
to be $(1.35\pm 0.51) \times 10^{-10}$ m s$^{-2}$. 
Using the methodology of our approach, constraining the gravitational potential
by proxy of the local distribution function, we found a similar value. If MOND is valid,
the simple interpolating function is more likely to be true to achieve the same
result from a global galactic-scale approach.

Adopting the standard interpolating function, the result of $a_0 \sim 1.2
    \times 10^{-10}$ m s$^{-2}$ is consistent with other studies employing MOND on
Galactic scales. This finding indicates that even in the solar neighborhood,
the prospect to observe the MOND effect is significant and compelling. Further
comprehensive investigations on this scale could shed light on the true nature
of MOND and dark matter.

\subsection{Comparisons}

Bayes factors (BF) serve as a tool for comparing two or more models based on
the observed data. The Bayes factor represents the integrated posterior
probability of a model across all potential parameter values, often referred
to as the ``evidence'' term in Bayesian statistics. A higher Bayes factor
indicates that a model is more suitable for explaining the observed data.
Basically, Bayes factor is the ratio of the likelihood of two hyphotheses.

\begin{figure}
    \includegraphics[width=\columnwidth]{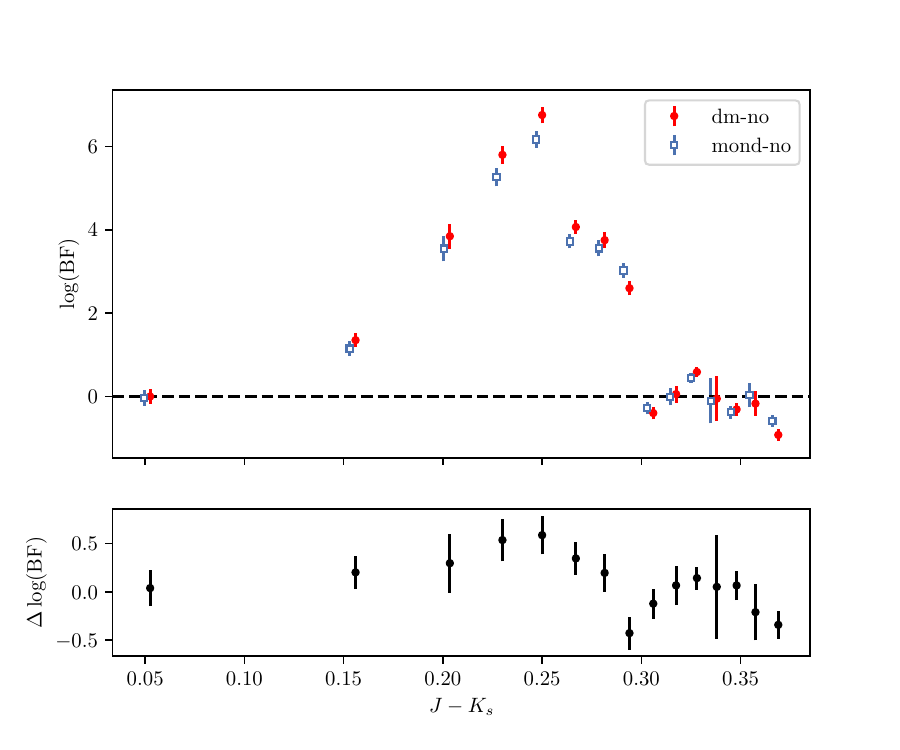}
    \caption{Bayes factors for each group of stars for the three hypotheses.
        (Top) Bayes factors of dark matter and MOND hypotheses with respect to NO hypothesis.
        The Bayes factor for MOND-NO points are slightly shifted to the left for clarity.
        (Bottom) Bayes factors comparing the dark matter hypothesis to the MOND
        hypothesis.}
    \label{fig:bayes-factors}
\end{figure}

The three hypotheses being compared are $M_{\textrm{no}}=$ Newtonian Dynamics \emph{with no} 
dark matter, $M_{\textrm{dm}}=$ Newtonian Dynamics \emph{with} dark matter, and $M_{\textrm{mond}}=$ MOND.
Figure \ref{fig:bayes-factors} presents the complete results.
This trend is more pronounced for stars of type F, where the value spikes to $6$.
Apart from these spikes, the $\log$ BF values are close to zero, indicating
that the three hypotheses are equally likely to be true. The only strong conclusion
that can be drawn is if the Newtonian dynamics is correct, then it must include 
non-baryonic dark matter, as indicated by the mean $\log$ BF values of
$M_{\textrm{dm}}$ with respect to $M_{\textrm{no}}$ for all groups of stars, which is
$1.80$ (dm-no). If we do not include the non-baryonic dark matter, MOND 
is more preferred, with mean $\log$ BF $1.70$ (mond-no).

Compare to the Bayes factor between the dark matter and MOND hypotheses; the two
are neck and neck. The $\log$ BF fluctuates around zero, with
the mean of only 0.10 in favor of dark matter. This result suggests that both
hypotheses are equally plausible in explaining the data we have. From our analysis,
we cannot definitively say which hypothesis is more favoured.

One way to better discriminate between the two hypotheses is by incorporating more 
data from stars at high altitudes ($|z| > 200$ pc). This approach aims to reveal 
potential differences in the predictions of the distribution function between the 
two hypotheses. However, a challenge arises for large altitudes as the assumption 
that the distribution function is separable becomes invalid. Consequently, solving 
the full form of the distribution function is necessary (see \eqref{eq:cbe}).
This poses a complex and challenging problem to address.

Other studies, such as the one conducted by Lisanti et al. (\citeyear{lisanti2019testing}),
have employed statistical methods to compare the dark matter and MOND hypotheses,
specifically by calculating the Bayesian Information Criterion (BIC). Their
findings strongly favour the dark matter hypothesis based on the difference in BIC
values ($\Delta \textup{BIC}$), which contrasts with our results. In our analysis, the
Bayes factors for the dark matter hypothesis are only marginally higher than those for the MOND
hypothesis. However, it is important to note that our results do not suggest
strong evidence against the MOND hypothesis, with the mean $\log$ BF values of $0.1$.
According to Kass \& Raftery (\citeyear{kass1995bayes}), a log Bayes factor of
less than 0.5 is not even worth more than a bare mention. More data and robust
methods are needed to provide conclusive answers.

\section{CONCLUSION} \label{sec:conclusion}

In this study, we have developed a method to estimate the dark matter density
in the solar neighbourhood by leveraging the vertical distribution of stars
obtained from Gaia, RAVE, APOGEE, GALAH, and LAMOST catalogues. The data collected
from these various catalogues are dominated by Gaia DR3. In principle, the analysis
done in this study is unaffected by the addition of other catalogues.

Our approach utilises both the vertical height distribution and vertical 
velocity distribution of stars as the vertical distribution function. In turn,
the distribution function is directly correlated with the Galactic gravitational
potential through CBE, which dictates the mass model for our hypotheses.
By proxy of the distribution function, the dark matter density and the acceleration 
parameter $a_0$ are inferred.

Our analysis reveals a dark matter density in the solar neighbourhood of
approximately $(0.01 \textrm{--} 0.03) \textrm{ M}_{\odot}$ pc$^{-3}$, and an acceleration
parameter $a_0$ of $1.26^{+0.14}_{-0.13} \times 10^{-10}$ m s$^{-2}$ for a simple
interpolating function. These fluctuations could be the results of disequilibria
of tracers being used, suggesting challenges to improve the approach for more robust 
results in future studies.

Bayesian analysis, through the calculation of Bayes factors, presents strong evidence
against classical Newtonian gravity without additional missing mass. However, when comparing
the two competing hypotheses in explaining the missing mass, we found practically no 
strong endorsement for either the dark matter hypothesis or the MOND hypothesis, with the Bayes 
factor between them being only $\log \textrm{BF} = 0.1$.
It is important to note, however, Bayes factors lower than $0.5$ provides no
substantial support for the dark matter hypothesis, falling short of
conclusive evidence to reject the MOND hypothesis outright.


\section*{Acknowledgements}

We thanked the Faculty of Mathematics and Natural Sciences of Bandung Institute of Technology
for financial support via the PPMI 2022 funding program.

This work has made use of data from the European Space Agency (ESA) mission
    {\it Gaia} (\url{https://www.cosmos.esa.int/gaia}), processed by the {\it Gaia}
Data Processing and Analysis Consortium (DPAC,
\url{https://www.cosmos.esa.int/web/gaia/dpac/consortium}). Funding for the DPAC
has been provided by national institutions, in particular the institutions
participating in the {\it Gaia} Multilateral Agreement.

This publication makes use of data products from the Two Micron All Sky Survey, which is a joint project of the
University of Massachusetts and the Infrared Processing and Analysis Center/California
Institute of Technology, funded by the National Aeronautics and Space Administration
and the National Science Foundation.

Funding for Rave has been provided by:
the Leibniz Institute for Astrophysics Potsdam (AIP);
the Australian Astronomical Observatory;
the Australian National University;
the Australian Research Council;
the French National Research Agency;
the German Research Foundation (SPP 1177 and SFB 881);
the European Research Council (ERC-StG 240271 Galactica);
the Istituto Nazionale di Astrofisica at Padova;
the Johns Hopkins University;
the National Science Foundation of the USA (AST-0908326);
the W. M. Keck foundation;
the Macquarie University;
the Netherlands Research School for Astronomy;
the Natural Sciences and Engineering Research Council of Canada;
the Slovenian Research Agency;
the Swiss National Science Foundation;
the Science \& Technology Facilities Council of the UK;
Opticon;
Strasbourg Observatory;
the Universities of Basel, Groningen, Heidelberg and Sydney.

This work has made use of data from the European Space Agency (ESA) mission Gaia
(\url{https://www.cosmos.esa.int/gaia}), processed by the Gaia Data Processing and
Analysis Consortium (DPAC, \url{https://www.cosmos.esa.int/web/gaia/dpac/consortium}).
Funding for the DPAC has been provided by national institutions, in particular
the institutions participating in the Gaia Multilateral Agreement.

Guoshoujing Telescope (the Large Sky Area Multi-Object Fiber Spectroscopic Telescope LAMOST) is a National Major Scientific Project built by the Chinese Academy of Sciences. Funding for the project has been provided by the National Development and Reform Commission. LAMOST is operated and managed by the National Astronomical Observatories, Chinese Academy of Sciences.

This work made use of the Third Data Release of the GALAH Survey 
(Buder et al. \citeyear{buder2021the}). The GALAH Survey is based on data acquired through the 
Australian Astronomical Observatory, under programs: A/2013B/13 (The GALAH pilot 
survey); A/2014A/25, A/2015A/19, A2017A/18 (The GALAH survey phase 1); A2018A/18 
(Open clusters with HERMES); A2019A/1 (Hierarchical star formation in Ori OB1); 
A2019A/15 (The GALAH survey phase 2); A/2015B/19, A/2016A/22, A/2016B/10, 
A/2017B/16, A/2018B/15 (The HERMES-TESS program); and A/2015A/3, A/2015B/1, 
A/2015B/19, A/2016A/22, A/2016B/12, A/2017A/14 (The HERMES K2-follow-up program). 
We acknowledge the traditional owners of the land on which the AAT stands, the 
Gamilaraay people, and pay our respects to elders past and present. This paper 
includes data that has been provided by AAO Data Central (\url{datacentral.org.au}).

Funding for the Sloan Digital Sky Survey V has been provided by the Alfred P. 
Sloan Foundation, the Heising-Simons Foundation, the National Science Foundation, 
and the Participating Institutions. SDSS acknowledges support and resources from 
the Center for High-Performance Computing at the University of Utah. The SDSS 
web site is \url{www.sdss.org}.

SDSS is managed by the Astrophysical Research Consortium for the Participating 
Institutions of the SDSS Collaboration, including the Carnegie Institution for 
Science, Chilean National Time Allocation Committee (CNTAC) ratified researchers, 
the Gotham Participation Group, Harvard University, Heidelberg University, 
The Johns Hopkins University, L'Ecole polytechnique f\'{e}d\'{e}rale de Lausanne
(EPFL), Leibniz-Institut f\"{u}r Astrophysik Potsdam (AIP), Max-Planck-Institut 
f\"{u}r Astronomie (MPIA Heidelberg), Max-Planck-Institut f\"{u}r 
Extraterrestrische Physik (MPE), Nanjing University, National Astronomical 
Observatories of China (NAOC), New Mexico State University, The Ohio State 
University, Pennsylvania State University, Smithsonian Astrophysical Observatory, 
Space Telescope Science Institute (STScI), the Stellar Astrophysics Participation 
Group, Universidad Nacional Aut\'{o}noma de M\'{e}xico, University of Arizona, 
University of Colorado Boulder, University of Illinois at Urbana-Champaign, 
University of Toronto, University of Utah, University of Virginia, 
Yale University, and Yunnan University.


\section*{Data Availability}
The original raw data are available from publicly availabe source at their respective
repositories. The reduced data underlying this article will be shared upon reasonable
request to the corresponding author.



\bibliographystyle{mnras}
\bibliography{references} 



\appendix
\section{Density Inference} \label{app:density-inference}

Let $x$ be an observed value with some uncertainty sampled from a distribution 
$\mathcal{T}$, with mean $\mu$ and standard deviation $\sigma$. The mean value 
$\mu$ can be inferred using Bayes' theorem:
\begin{align}
    p(\mu|x,\sigma) & = \frac{p(x|\mu,\sigma)p_0(\mu|\sigma)}{E(x,\sigma)} \nonumber \\
                    & \propto \mathcal{T}(x|\mu,\sigma),  \label{eq:inference}
\end{align}
where $p_0(\mu|\sigma)\propto 1$ is assumed. If there is a set of observed 
values $\{\bm{x}, \bm \sigma\}$ that are independent and identically distributed 
(iid), the probability density function of the mean value $\bm \mu$ is
\begin{align}
    p(\bm \mu|\bm x, \bm \sigma) & \propto \prod_{i=1}^{N} \mathcal{T}(x_i|\mu_i,\sigma_i). \label{eq:inference-many}
\end{align}

Consider an interval $x$ and $x+\Delta x$. Due to the noisy nature of the data, 
the number of observed values in that interval is a random variable. Let $N$ 
be the number of data points with mean $\mu$ value within the interval. The 
probability density function of $N$ is given by
\begin{align}
    p(N|\bm x, \bm \sigma) &= \int p(N, \bm \mu|\bm x, \bm \sigma) \textrm{d}\bm \mu \nonumber\\
    &= \int p(N|\bm \mu, \bm x, \bm \sigma)p(\bm \mu|\bm x, \bm \sigma) \textrm{d}\bm \mu \nonumber\\
    &= \int p(N|\bm \mu) p(\bm \mu|\bm x, \bm \sigma) \textrm{d}\bm \mu \nonumber\\
    &= \int \mathbb{I}(\bm\mu)p(\bm \mu|\bm x, \bm \sigma) \textrm{d}\bm \mu, \label{eq:inference-count}
\end{align}
where $\mathbb{I}(\bm \mu)$ is one when the combination of $\bm \mu$ results in $N$ 
observed values in the interval, and zero otherwise. In reality, the random 
variable $\mu$ is distributed according to a Poisson distribution. Let 
$\Lambda$ be the expected number of observed values of $\mu$ in the interval; 
then, the probability density function of $\Lambda$ is given by
\begin{align}
    p(\Lambda|\bm x, \bm \sigma) &= \sum p(\Lambda, N_i| \bm x, \bm \sigma) \nonumber\\
    &= \sum p(\Lambda|N_i, \bm x, \bm \sigma)p(N_i|\bm x, \bm \sigma) \nonumber\\
    &= \sum p(\Lambda|N_i)\int \mathbb{I}(\bm \mu) p(\bm \mu|\bm x, \bm \sigma) \textrm{d}\bm \mu \nonumber\\ 
    &= \sum W(N_i)p(\Lambda|N_i), \label{eq:inference-poisson}
\end{align}
where $p(\Lambda|N_i)\propto p(N_i|\Lambda)$ is the probability density function 
of the Poisson distribution with mean $\Lambda$ given an uninformative prior, and 
$W(N_i)$ is the weight of finding $N_i$ observed values in the interval. The 
weight $W(N_i)$ is extremely hard to calculate from the integral in equation 
\eqref{eq:inference-count}, so in practice
we use the following approximation:
\begin{align}
    W(N_i) \approx \frac{1}{S}\sum_{j=1}^S \mathbb{I}(\bm \mu_j), \label{eq:weight}
\end{align}
where we sample $\bm \mu$ from $p(\bm \mu|\bm x, \bm \sigma)$ $S$ times.

Equation \eqref{eq:inference-poisson} is computationally inefficient due to the 
large number of possible values for $N_i$. To address this, we defined a new
distribution function called the \emph{modified Poisson distribution} to 
approximate the probability density function of $\Lambda$:
\begin{align}
    \textrm{Poisson}^*(N|\Lambda) = \frac{\Lambda^Ne^{-\Lambda}}{\Gamma (N+1)}, \label{eq:modified-poisson} 
\end{align}
where $\Gamma(x)$ is the gamma function. Our tests show that the modified Poisson 
distribution is a good approximation for equation \eqref{eq:inference-poisson}. 
Thus, we express the probability density function of $\Lambda$ as
\begin{align}
    p(\Lambda|\bm x, \bm \sigma) & \approx \textrm{Poisson}^*(N_{\textrm{eff}}|\Lambda), \label{eq:inference-poisson-approx}
\end{align}
where $N_{\textrm{eff}}$ represents the \emph{effective number of observed values} 
in the interval. This number is not necessarily an integer. This choice minimizes the mean square error between equation 
\eqref{eq:inference-poisson} and equation \eqref{eq:inference-poisson-approx}, 
ensuring both computational efficiency and a close approximation to the true value.


\bsp	
\label{lastpage}
\end{document}